\documentclass[aps,prd,preprint,showpacs,groupedaddress]{revtex4}
\usepackage{amsmath}
\usepackage{tipa}
\usepackage{bbm}
\usepackage{txfonts}
\usepackage{graphicx}   
\usepackage{dcolumn}    
\usepackage{bm}         
\usepackage{amssymb}    
\usepackage{latexsym}   

\begin{document}

\preprint{Submitted to PRL}

\title{Raman pulse duration effect in gravity gradiometers composed of two atom interferometers}
\author{Cheng-Gang Shao}
\author{De-Kai Mao}

\author{Min-Kang Zhou}
\author{Yu-Jie Tan}
\author{Le-Le Chen}
\author{Jun Luo}
\author{ZhongKun-Hu}
\email[Corresponding author: ]{zkhu @ mail.hust.edu.cn}

\affiliation{Key Laboratory of Fundamental Physical Quantities
Measurement of Ministry of Education, School of Physics, Huazhong
University of Science and Technology, Wuhan 430074, People's
Republic of China}


\begin{abstract}
We investigated the Raman pulse duration effect in a gravity gradiometer with two atom interferometers. Since the two atom clouds in the gradiometer experience different gravitational fields, it is hard to compensate the Doppler shifts of the two clouds simultaneously by chirping the frequency of a common Raman laser, which leads to an appreciable phase shift. When applied to an experiment measuring the Newtonian gravitational constant G , the effect contributes to a systematic offset as large as -49ppm in Nature 510, 518 (2014). Thus an underestimated value of G measured by atom interferometers can be partly explained due to this effect.
\end{abstract}


\maketitle Atom interferometry, with great sensitivity to accelerations, is widely used in precision measurements \cite{1,2,3,4,5,6,7,8}. Gravity gradiometers composed of two atom interferometers have been employed to measure the gravity gradient\cite{4,5,6,8,9} and the Newtonian gravitational constant  \cite{1,3,5,9,10,11}. Three conjugated atom interferometers is also presented to directly measure the gravity-field curvature \cite{2}. In these high precision measurements, the Raman pulses interrogate two or more separated atom clouds simultaneously in order to reject the common-mode vibration noise. The frequency of the Raman pulses sweeps linearly with time to compensate the doppler effect. But the compensation is not adequate for all the conjugated interferometers since they generally experience different gravitational fields. We find this inadequate compensation contributes to a quite large systematic offset, and can not be ignored. We focus on the gravity gradiometer with two atom interferometers in this paper.

The effect of the Raman pulse duration effect in an atom interferometer experiencing a homogeneous gravitational field has been studied in  \cite{13,14,15,16,17}. The phase shift of the interferometer can be written as follows,
\begin{equation}
\Delta \varphi=-(\alpha+\overrightarrow{k}\cdot
\overrightarrow{g})T^2\left(1-\frac{2\tau}{T}+\frac{4\tau
}{\pi T}\right)
,\label{Eq:1}
\end{equation}
Where $\alpha$ is the chirp rate of the light frequency with time, $\overrightarrow{k}$ is the effective wave vector of the Raman laser,$\overrightarrow{g}$ is the gravitational acceleration, $\tau$ is the duration of a $\pi/2 $ pulse, and $T $ is the interval time between two Raman pulses. Formula (\ref{Eq:1}) is equivalent to that in \cite{7,8,9,10,11} considering the different definition of $T $ we used. In this paper, $T $ lasts from the beginning of the first pulse to the middle of the second pulse (see FIG. 1).

For the central fringe, we have $\alpha+\overrightarrow{k}\cdot\overrightarrow{g}\approx 0$, then the local gravity  $\overrightarrow{g} $ is determined by $\alpha $  and $\overrightarrow{k} $ . The Raman pulse duration $\tau$ only causes negligible offset in an atom gravimeter.

In a gravity gradiometer composed of two atom interferometers, the gravity field that the two atom clouds experience is denoted as $\overrightarrow{g}^{up} $ and  $\overrightarrow{g}^{low} $ in respective heights, where the superscripts  $up $ and $low $  indicate the upper and lower interferometers. Noting that the laser phase shared by the two clouds is canceled, the differential phase shift of the two atom interferometers can be written as
\begin{equation}
\Delta \varphi^{up}-\Delta \varphi^{low}=-\overrightarrow{k}\cdot
(\overrightarrow{g}^{up}-\overrightarrow{g}^{low})T^2\left(1-\frac{2\tau}{T}+\frac{4\tau
}{\pi T}\right)
.\label{Eq:2}
\end{equation}
This formula shows that the differential phase shift is associated with the Raman pulse duration. The relative offset of the differential phase shift caused by the Raman pulse duration effect is evaluated at $-0.73\tau/T$.

\begin{figure}[tbp]
\includegraphics[trim=40 10 40 20,width=0.40\textwidth]{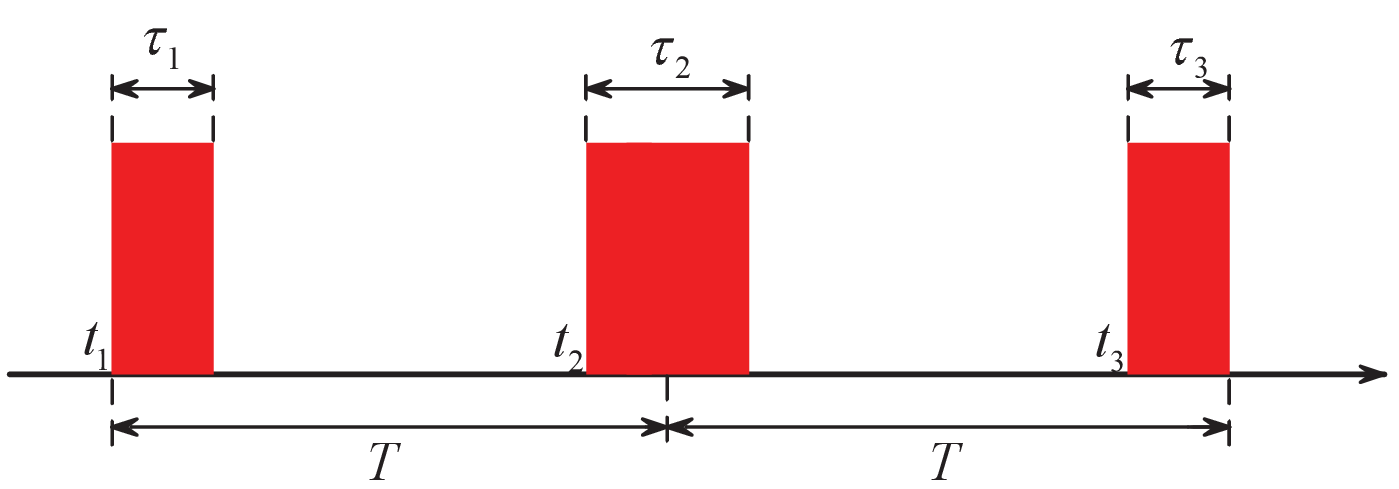}
\caption{\label{fig:1}(color online). Diagram of the time sequence of a three-pulse Mach-Zehnder type interferometer.}
\end{figure}

In the experiment measuring the Newtonian gravitational constant G  \cite{1}, the position of the Source Mass is modulated between the far and close configurations to cancel the influence of the background gravitational field. The differential phase shift is labeled as $\phi_C $  and $\phi_F $ corresponding to the close and far configurations. Since the Newtonian gravitational constant G  is proportional to $\phi_C-\phi_F $  in the experiment, G is extracted by comparing the simulated and measured differential phase shifts. In the homogenous gravitational field approximation, the phase $\phi_C-\phi_F $  should be estimated as
\begin{equation}
\begin{split}
\phi_C-\phi_F=&
-\overrightarrow{k}\cdot
(\overrightarrow{a}_{C}^{up}-\overrightarrow{a}_{C}^{low})
T^2 \left(1-\frac{2\tau}{T}+\frac{4\tau
}{\pi T}\right) \\
&+\overrightarrow{k}\cdot
(\overrightarrow{a}_{F}^{up}-\overrightarrow{a}_{F}^{low})
T^2 \left(1-\frac{2\tau}{T}+\frac{4\tau
}{\pi T}\right)
\end{split}
\label{Eq:3}
\end{equation}
where $\overrightarrow{a}$ indicates the average gravitational acceleration each interferometer experiences respectively (see FIG. 2). The authors of \cite{1} use a perturbation treatment of Feynman path integration \cite{11,22} in the experiment. Comparing with the phase shift $-\overrightarrow{k}\cdot
(\overrightarrow{a}_{C}^{up}-\overrightarrow{a}_{C}^{low})T^2 +\overrightarrow{k}\cdot
(\overrightarrow{a}_{F}^{up}-\overrightarrow{a}_{F}^{low})T^2$ obtained from their treatment, a -54.5ppm systematic offset appeared when the relative parameters $\tau $  ($12 \mu s $ ) and  $T $ ( $160 ms $) of the experiment are substituted in. So in this approximation, the G value in \cite{1} will be 54.5 ppm larger if they had considered this offset.

\begin{figure}[tbp]
\includegraphics[trim=40 10 40 20,width=0.45\textwidth]{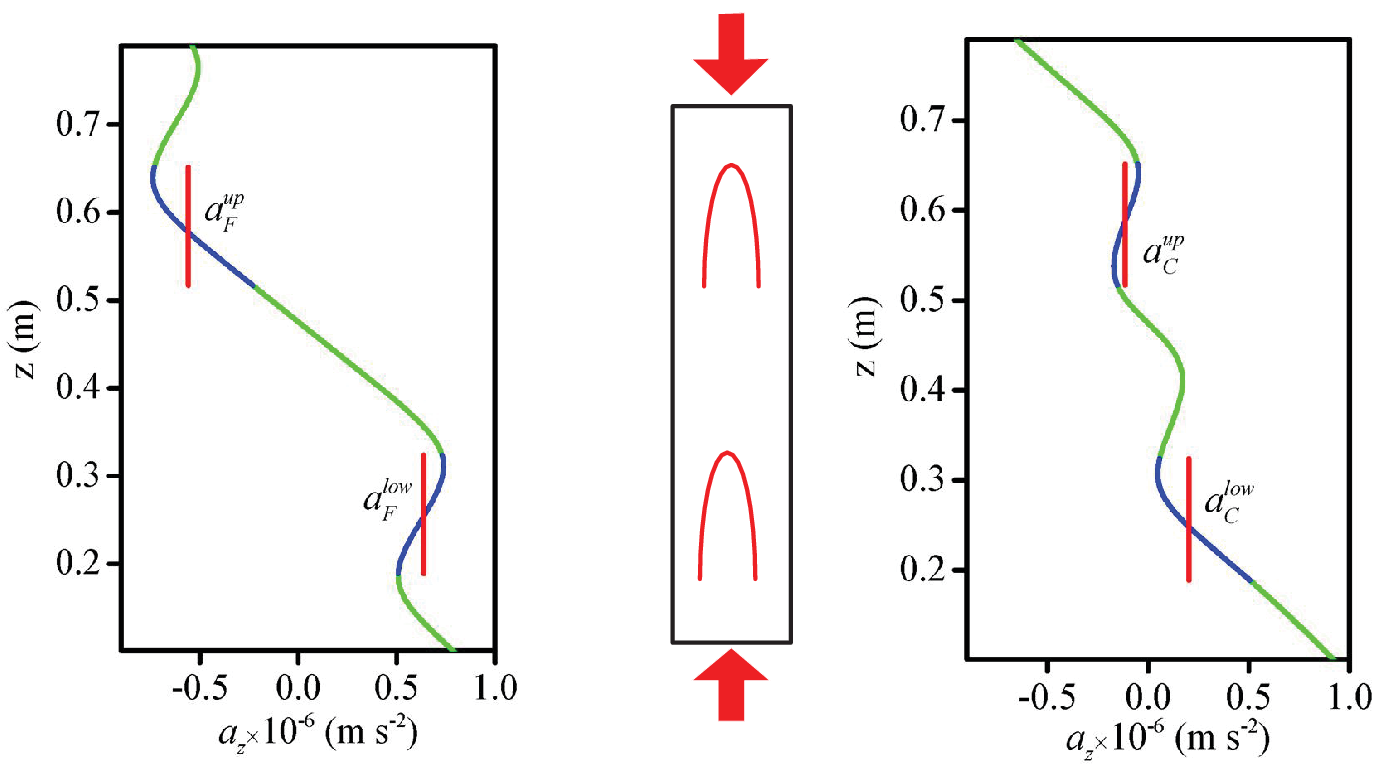}
\caption{\label{fig:2}(color online). The gravitational field along the vertical symmetry axis of the Far (left) and Close (right) configuration (with data from \cite{1,11,19,20,21}). The thick lines on the plots indicate the atom interferometers' spatial regions. The vertical bar across a thick line represents the mean value over the atom interferometer's spatial region.}
\end{figure}

In the rest of this paper, a more sophisticated treatment is performed considering the actual inhomogeneous gravitational field (FIG.2). We have already developed a method to derive the phase shift of an interferometer in an inhomogeneous gravitational field \cite{18}. The phase shift of an interferometer is calculated based on this method to track the systematic error in the G measuring experiment.

As shown in FIG. 1, the start point of time is set to be at $t_1 $  (namely $t_1=0 $ ) to simplify the calculation. After multiplying the evolution operators along the interference sequence, the expression of the phase shift is obtained. The total phase shift is divided into two parts,
\begin{equation}
\Delta\phi_{total}=\Delta\phi_{laser}+\Delta\phi_{prop}.
\label{Eq:4}
\end{equation}
$\Delta\phi_{laser} $  is the phase shift from the atom interaction with the laser,
\begin{equation}
\Delta\phi_{laser}=\left(\phi_1+\frac{\delta_1\tau_1}{2}\left(2-\frac{4}{\pi}\right)\right)
-2\left(\phi_2+\frac{\delta_2\tau_2}{2}\right)
+\left(\phi_3+\frac{\delta_3\tau_3}{2}\frac{4}{\pi}\right).
\label{Eq:5}
\end{equation}
With a linear variation of laser frequency with time, we get
\begin{equation}
\phi_1-2\phi_2+\phi_3=\omega_L(t_3-2t_2)-\frac{\alpha
}{2}(t_3^2-2t_2^2)
\label{Eq:6}
\end{equation}
and
\begin{equation}
\begin{split}
&\delta_1=\delta_0 \\
&\delta_2=\delta_0-(\alpha+\overrightarrow{k}\cdot\overrightarrow{g})t_2
-\overrightarrow{k}\cdot\int_{0}^{t_2}\overrightarrow{a}(t)dt\\
&\delta_3=\delta_0-(\alpha+\overrightarrow{k}\cdot\overrightarrow{g})t_3
-\overrightarrow{k}\cdot\int_{0}^{t_3}\overrightarrow{a}(t)dt\\
\end{split}.
\label{Eq:7}
\end{equation}
$\Delta\phi_{prop} $represents the phase shift from the free evolution,
\begin{equation}
\begin{split}
\Delta\phi_{prop}&=\omega_{eg}(2t_2-t_3)+\overrightarrow{k}\cdot
\overrightarrow{v}_0(2t_2-t_3)+\frac{1}{2}\overrightarrow{k}\cdot
\overrightarrow{g}\left(2t_2^2-t_3^2\right)\\
&-\overrightarrow{k}\cdot\left[\int_{0}^{t_2}(t+t_3-2t_2)\overrightarrow{a}(t)dt+
\int_{t_2}^{t_3}(t_3-t)\overrightarrow{a}(t)dt\right]
\end{split}.
\label{Eq:8}
\end{equation}
Here $\phi_i $  and  $\delta_i $ ( $i=1,2,3 $) denote the laser phase and the detuning of the laser frequency at the time  $t_i $,  $\tau_i $ is the  $i $th Raman pulse duration time,  $\overrightarrow {v}_0 $ is the average velocity of an atom after the first $\pi/2 $  Raman pulse, $\omega_L $  is the laser frequency at $t_1 $ , $\omega_{eg} $  represents the transition frequency of a two-level atom between ground and excited states.

$\overrightarrow{a}(t)$ comes from $\overrightarrow{a}(\overrightarrow{z})$, which describes the gravity along the vertical axis caused by the source mass and the Earth's gravitational gradient. When an atom moved to $\overrightarrow{z} $  position in its classical path at time $t $ , the acceleration it is experiencing is $\overrightarrow{g}+\overrightarrow{a}(\overrightarrow{z}(t)) $ . Noting that  $|\overrightarrow{a}(\overrightarrow{z}(t))|/|\overrightarrow{g}|\approx 10^{-7} $, the classical path of atom is mainly determined by Earth's gravity $|\overrightarrow{g}|=9.8056 m/s^2 $   \cite{23}. So $\overrightarrow{z}(t) $  is expressed as
\begin{equation}
\overrightarrow{z}(t)=\overrightarrow{z}_0+\overrightarrow{v}_0 t
+\frac{1}{2}\overrightarrow{g}t^2.
\label{Eq:9}
\end{equation}

According to the interference sequence of the interferometer used in the experiment \cite{20}, the time parameter is determined as follows,
\begin{equation}
\begin{split}
&\tau_1=\tau_3=\tau,\tau_2=2\tau\\
&t_2=T-\tau,t_3=2T-\tau
\end{split}.
\label{Eq:10}
\end{equation}
From (\ref{Eq:4}) to (\ref{Eq:10}), ignoring all the terms higher than first order of $\tau $ , the total phase shift of one atom interferometer is given as
\begin{equation}
\begin{split}
\Delta \varphi_{total}&=-(\alpha+\overrightarrow{k}\cdot \overrightarrow{g})
T^2\left(1-\frac{2\tau}{T}+\frac{4\tau}{\pi T}\right) \\
&-\overrightarrow{k}\cdot\left(\int_{0}^{T}t\overrightarrow{a}(t)dt+\int_{T}^{2T}(2T-t)\overrightarrow{a}(t)dt\right)\\
&-2\tau \overrightarrow{k}\cdot\left(\frac{1}{\pi}\int_{0}^{2T}\overrightarrow{a}(t)dt-\int_{0}^{T}\overrightarrow{a}(t)dt\right) \\
&+\tau \overrightarrow{k}\cdot\left[\int_{0}^{T}t(\overrightarrow{v_0}+\overrightarrow{g}t)\frac{d\overrightarrow{a}}{d\overrightarrow{z}}dt
+\int_{T}^{2T}(2T-t)(\overrightarrow{v_0}+\overrightarrow{g}t)\frac{d\overrightarrow{a}}{d\overrightarrow{z}}dt\right]
\end{split}.
\label{Eq:11}
\end{equation}

 In this formula, the total phase shift is the sum of four terms. Since we choose the same $\overrightarrow{g} $  for the two atom clouds, the first term of $\Delta \phi_{total} $  is cancelled in  $\phi_C $ and $\phi_F $ . The second term of  $\Delta \phi_{total} $ equals to the perturbation treatment of a Feynman path integration\cite{22}, which is shown as
\begin{equation}
\begin{split}
\phi_{pert}&=-\frac{1}{\hslash}\int_{0}^{2T}(V_{ACB}-V_{ADB})dt\\
&=-\frac{1}{\hslash}\int_{0}^{2T}\frac{dV}{d\overrightarrow{z}}\Delta \overrightarrow{z}dt\\
&=-\frac{1}{\hslash}\int_{0}^{T}m \overrightarrow{a}(t)\frac{\hslash \overrightarrow{k}}{m}tdt
-\frac{1}{\hslash}\int_{T}^{2T}m \overrightarrow{a}(t)\frac{\hslash \overrightarrow{k}}{m}(2T-t)dt\\
&=-\overrightarrow{k}\cdot\left(\int_{0}^{T}t\overrightarrow{a}(t)dt+\int_{T}^{2T}(2T-t)\overrightarrow{a}(t)dt\right)
\end{split},
\label{Eq:12}
\end{equation}
Where $V $  is the perturbation of atom's Lagrangian. The distance $\Delta z $  between the two arms of the interferometer is shown in FIG.3. The third and the fourth terms of formula (\ref{Eq:11}) are new terms with relation to the first order of pulse duration time $\tau $ . Then the phase $\phi_C-\phi_F $  is expressed as
\begin{equation}
\phi_C-\phi_F =(\Delta\phi_{total,C}^{up}-\Delta\phi_{total,C}^{low})-(C\longleftrightarrow F).
\label{Eq:13}
\end{equation}

\begin{figure}[tbp]
\includegraphics[width=0.4\textwidth]{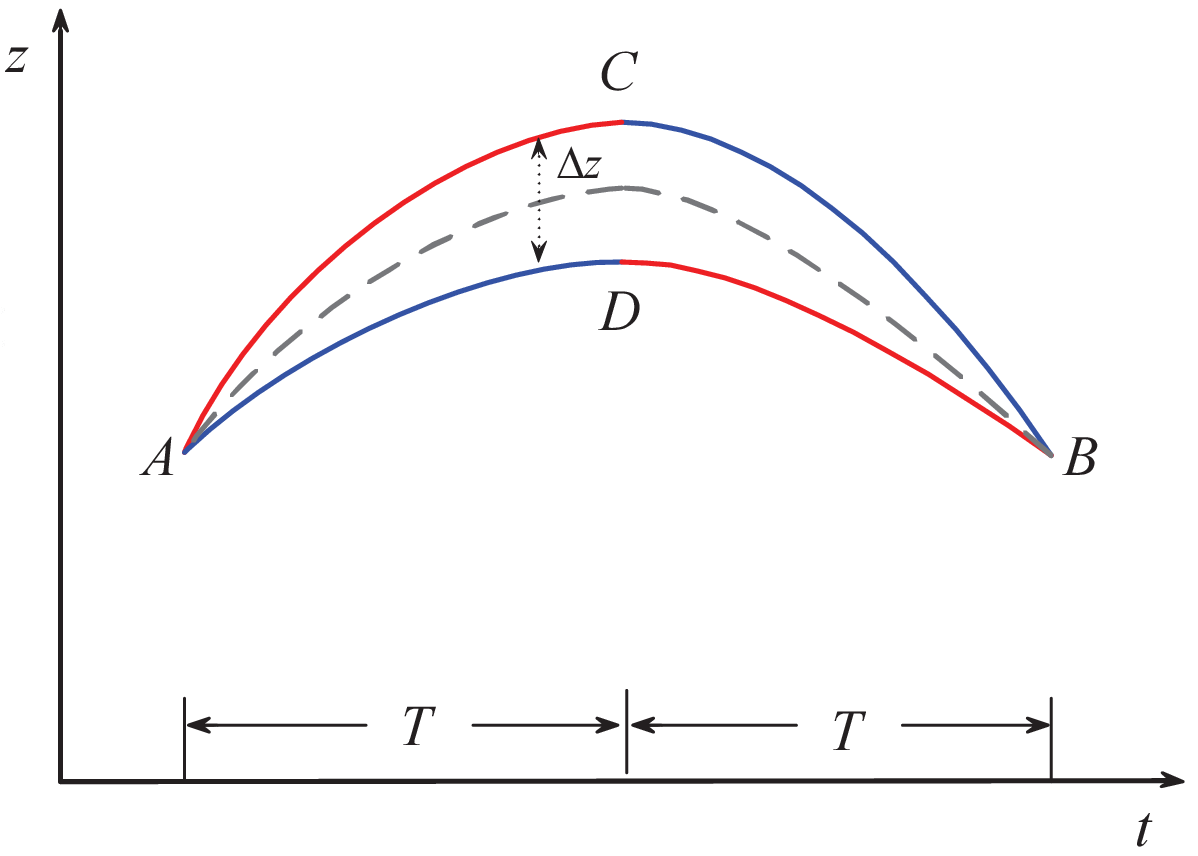}
\caption{\label{fig:3}(color online). Space-time diagram of an atom interferometer.  $\Delta z $ is the spacial distance between the two arms and it varies linearly with time.}
\end{figure}

Utilizing the gravitational field data shown in FIG.2, we compare   $\phi_C-\phi_F $ with the same quantity generated only from a simple perturbation treatment of Feynman path integration (the second term in formula(\ref{Eq:11})), about a -49 ppm systematic offset is obtained. This offset is fairly big compared with the total systematic uncertainty 92 ppm in the G measurement experiment. And it leads to a change of the G value from $6.67191(99)\times10^{-11} m^3kg^{-1}s^{-2} $ to $6.67224(99)\times10^{-11} m^3kg^{-1}s^{-2} $ .

A similar result can also be obtained with the sensitivity function of an atomic gravimeter \cite{24,25}. According to the definition of the sensitivity function, the interferometric phase shift $\delta \phi_{sens} $  for a Raman phase noise caused by  $\overrightarrow{v}(t) $ is evaluated as
\begin{equation}
\delta\phi_{sens}=\overrightarrow{k}\cdot\int_{0}^{2T}
g_s(t)\overrightarrow{v}(t)dt,
\label{Eq:14}
\end{equation}
where $\overrightarrow{v}(t) $  is the mean velocity of an atom wave package. In the presence of gravity, an atom is accelerated to move. Once having a displacement, it has an instantaneous phase difference, which is equivalent to a phase jump performed by the Raman laser. The velocity of atoms at time  $t $ is
\begin{equation}
\overrightarrow{v}(t)=\overrightarrow{v}_0+\overrightarrow{g}
t+\int_{0}^{t}\overrightarrow{a}(t')dt'.
\label{Eq:15}
\end{equation}
$g_s(t) $ is the sensitivity function. With our definition of time in the interference sequence, it is written as
\begin{equation}
g_s(t) = \begin{cases}

  sin(\Omega_rt) & \text{ $0\le t<\tau $} \\

  1 & \text{ $\tau\le t<T-\tau $} \\

  -sin[\Omega_r( t-T)] & \text{ $ T-\tau\le t<T+\tau $} \\

  -1 & \text{ $T+\tau\le t<2T-\tau $} \\

  -sin[\Omega_r( 2T-t)] & \text{ $ 2T-\tau\le t<2T $}
\end{cases}.
\label{Eq:16}
\end{equation}
With (\ref{Eq:15}) and (\ref{Eq:16}), the total phase shift derived from the sensitivity function of an atom interferometer becomes
\begin{equation}
\begin{split}
\Delta \varphi_{sens}&=-(\alpha+\overrightarrow{k}\cdot \overrightarrow{g})
T^2\left(1-\frac{2\tau}{T}+\frac{4\tau}{\pi T}\right) \\
&-\overrightarrow{k}\cdot\left(\int_{0}^{T}t\overrightarrow{a}(t)dt+\int_{T}^{2T}(2T-t)\overrightarrow{a}(t)dt\right)\\
&-2\tau \overrightarrow{k}\cdot\left(\frac{1}{\pi}\int_{0}^{2T}\overrightarrow{a}(t)dt-\frac{1}{2}\int_{0}^{2T}\overrightarrow{a}(t)dt\right)
\end{split}.
\label{Eq:17}
\end{equation}
Here, we have just kept to the first order term of $\tau $ , and a linear frequency ramp $\alpha $  has been added to the Raman laser to preserve the resonance condition.

We find that the difference between $\Delta \phi_{total} $  and  $\Delta \phi_{sens} $ vanishes identically by a calculation using an integration by parts.
\begin{equation}
\begin{split}
&\Delta \varphi_{total}-\Delta \varphi_{sens}=\\
&\tau \overrightarrow{k}\cdot\left[\int_{0}^{T}t(\overrightarrow{v_0}+\overrightarrow{g}t)\frac{d\overrightarrow{a}}{d\overrightarrow{z}}dt
+\int_{T}^{2T}(2T-t)(\overrightarrow{v_0}+\overrightarrow{g}t)\frac{d\overrightarrow{a}}{d\overrightarrow{z}}dt\right]\\
&+\tau \overrightarrow{k}\cdot\left(\int_{0}^{T}\overrightarrow{a}(t)dt-\int_{T}^{2T}\overrightarrow{a}(t)dt\right)
\equiv 0
\end{split}.
\label{Eq:18}
\end{equation}
Namely, the method arising from the sensitivity function is equivalent to the method using the time evolution operators considering only to the first order terms of $\tau$. But to the best of our knowledge, the strictness of the phase shift calculation utilizing a sensitivity function has not been proved previously.

In conclusion, we report a systematic effect in a gravity gradiometer composed of two atom interferometers which has not been included by previous works. We derived the phase shift of an atom interferometer under an arbitrary gravity field distribution considering the Raman pulse duration effect. Evaluating the Raman pulse effect in the experiment measuring the gravitational constant G, we find a -49 ppm systematic offset with the typical experiment parameters.

We gratefully acknowledge support from the National Natural Science Foundation of China (Grant Nos. 11275075, 10875045).

\cleardoublepage


\begin{thebibliography}{99}
\bibitem{1} G. Rosi, F. Sorrentino, L. Cacciapuoti, M. Prevedelli, and G. M. Tino, Nature 510, 518 (2014).
\bibitem{2} G. Rosi, L. Cacciapuoti, F. Sorrentino, M. Menchetti, M. Prevedelli, and G. M. Tino, Phys. Rev. Lett. 114, 013001 (2015).
\bibitem{3} J. Fixler, G. Foster, J. McGuirk, and M. A. Kasevich, Science 315, 74 (2007).
\bibitem{4} J. M. McGuirk, G. T. Foster, J. B. Fixler, M. J. Snadden, and M. A. Kasevich, Phys. Rev. A 65, 033608 (2002).
\bibitem{5} G. W. Biedermann, X. Wu, L. Deslauriers, S. Roy, C. Mahadeswaraswamy, and M. A. Kasevich, arXiv 1412.3210v1 (2014).
\bibitem{6} X. C. Duan, M. K. Zhou, D. K. Mao, H. B. Yao, X. B. Deng, J. Luo, and Z. K. Hu, Phys. Rev. A 90,  023617 (2014).
\bibitem{7} Z. K. Hu, B. L. Sun, X. C. Duan, M. K. Zhou, L. L. Chen, S. Zhan, Q. Zhang, and J. Luo, Phys. Rev. A 88, 043610 (2013).
\bibitem{8} M. K. Zhou, Z. K. Hu, X. C. Duan, B. L. Sun, L. L. Chen, Q. Z. Zhang, and J. Luo, Phys. Rev. A 86, 043630 (2012).
\bibitem{9} N. Yu, J. M. Kohel, J. R. Kellogg, and L. Maleki, Appl. Phys. B  84, 647 (2006).
\bibitem{10} F. Sorrentino, Q. Bodart, L. Cacciapuoti, Y. Lien, M. Prevedelli, G. Rosi, L. Salvi, and G. Tino, Phys. Rev. A 89, 023607 (2014).
\bibitem{11} M. Prevedelli, L. Cacciapuoti, G. Rosi, F. Sorrentino, and G. M. Tino, Phil. Trans. R. Soc. A 372, 20140030 (2014).
\bibitem{12} G. Lamporesi, A. Bertoldi, L. Cacciapuoti, M. Prevedelli, and G. M. Tino, Phys. Rev. Lett. 100, 050801 (2008).
\bibitem{13} A. Peters, Ph.D. Thesis (Stanford University) (1998).
\bibitem{14} C. Antoine, Appl. Phys. B 84, 585 (2006).
\bibitem{15} C. Antoine, Phys. Rev. A 76, 033609 (2007).
\bibitem{16} M. A. H. M. Jansen and K. A. H. van Leeuwen, Appl. Phys. B 93, 389 (2008).
\bibitem{17} R. Stoner, D. Butts, J. Kinast, and B. Timmons, J. Opt. Soc. Am. B 28, 2418 (2011).
\bibitem{18} X. Li, C. G. Shao, and Z. K. Hu, J. Opt. Soc. Am. B 32, 248 (2015).
\bibitem{19} G. Lamporesi, A. Bertoldi, A. Cecchetti, B. Duhlach, M. Fattori, A. Malengo, S. Pettorruso, M. Prevedelli, and G. Tino, Rev. Sci. Instrum 78, 075109 (2007).
\bibitem{20} G. Lamporesi, Ph.D. Thesis, University of Florence (2006).
\bibitem{21} G. Rosi, Ph.D. Thesis, University of Florence (2012).
\bibitem{22} P. Storey and C. Cohen-Tannoudji, J. Phys. II  4, 1999 (1994).
\bibitem{23} A. Bertoldi, G. Lamporesi, L. Cacciapuoti, M. Angelis, M. Fattori, T. Petelski, A. Peters, M. Prevedelli, J. Stuhler, and G.M. Tino, Eur. Phys. J. D 40, 271279 (2006).
\bibitem{24} P. Cheinet, B. Canuel, F. Santos, A. Gauguet, F. Yver-Leduc, and A. Landragin, IEEE Trans. Instrum. Meas. 57, 11411148 (2008).
\bibitem{25} P. Lemonde, G. Santarelli, P. Laurent, F. Santos, A. Clairon, and C. Salomon, IEEE Proc. Frequency Control Symp. 110 (1998).



\end{thebibliography}
\end{document}